\documentclass[12pt]{article}
\usepackage{amssymb, amsmath, bm, float, epsfig, color}

\newcommand{\eq}[1]{Eq.~(\ref{#1})}

\newcommand{\rf}[1]{Ref.~\cite{#1}}
\newcommand{\rfs}[1]{Refs.~\cite{#1}}

\begin{document}
\title{
\vspace{-20pt}
\begin{flushright}
\normalsize WU-HEP-16-20 \\*[55pt]
\end{flushright}
{\Large \bf 
Inflation from periodic extra dimensions
\\*[20pt]}
}

\author{
Tetsutaro~Higaki,$^1$\footnote{E-mail address: thigaki@rk.phys.keio.ac.jp} \, and \,
Yoshiyuki~Tatsuta$^2$\footnote{E-mail address: y\_tatsuta@akane.waseda.jp}\\*[30pt]
$^1${\it \normalsize Department of Physics, Keio University, Kanagawa 223-8522, Japan}\\
$^2${\it \normalsize Department of Physics, Waseda University, Tokyo 169-8555, Japan}\\*[55pt]
}

\date{
\centerline{\small \bf Abstract}
\begin{minipage}{0.9\textwidth}
\medskip\medskip 
\small
We discuss a realization of a small field inflation 
based on string inspired supergravities.
In theories accompanying extra dimensions, compactification of them with small radii is required for realistic situations.
Since the extra dimension can have a periodicity, 
there will appear (quasi-)periodic functions under transformations of moduli of
the extra dimensions in low energy scales. 
Such a periodic property can lead to a UV completion of so-called multi-natural inflation model where inflaton potential consists of
a sum of multiple sinusoidal functions with a decay constant smaller than the Planck scale.
As an illustration, we construct a SUSY breaking model, 
and then show that such an inflaton potential can be generated 
by a sum of world sheet instantons 
in intersecting brane models on extra dimensions containing 
$T^2/{\mathbb Z}_2$
orbifold.
We show also predictions of cosmic observables by numerical analyzes.
%
\end{minipage}}

\begin{titlepage}
\maketitle
\thispagestyle{empty}
\clearpage
\tableofcontents
\thispagestyle{empty}
\end{titlepage}

\section{Introduction}
The cosmic microwave background (CMB) observation excellently supports the presence of cosmic inflation, 
which is
an exponential expansion of space in the early universe 
\cite{Guth:1980zm, Sato:1980yn, Brout:1977ix, Kazanas:1980tx, Starobinsky:1980te}.
The rapid expansion of the inflation scenario can solve the cosmological problems associated with two fine-tunings in the standard cosmology, i.e., the horizon problem and the flatness problem, simultaneously. Further, the fluctuation of the CMB temperature 
is explained by that of an inflaton scalar field, which drives the inflation.  
The scenario has been steadily tested in recent cosmic observations.
In particular, slow-roll inflation scenario agrees with
recent precise observations.
The Planck observation shows the values of the spectral index and its running \cite{Planck:2013jfk, Ade:2015lrj},
\begin{gather}
n_s = 0.9655 \pm 0.0062 \quad (68\%~{\rm CL}),\\
\alpha_s = \frac{dn_s}{d \ln k} = - 0.0033 \pm 0.0074 \quad (68\%~{\rm CL}).
\end{gather}
The observation also constrains the tensor-to-scalar ratio as
\begin{gather}
r < 0.10 \quad (95\%~{\rm CL}). 
\label{planck}
\end{gather}
These imply that the fluctuation of the CMB is almost scale invariant and the
primordial gravitational wave caused by
the inflation is not found yet at the current stage.
Then, inflation models have been strictly classified by these facts.

An approximate shift symmetry of the inflaton
\begin{gather}
\phi \to \phi + 2\pi f,
\end{gather}
often plays important roles
in controlling the flatness of the scalar potential
\cite{Freese:1990rb,Kawasaki:2000yn,Silverstein:2008sg,McAllister:2008hb,Kaloper:2008fb}.
Here, $f$ is a decay constant of an inflaton field $\phi$. 
Natural inflation \cite{Freese:1990rb} has often been considered so far,
and the inflaton potential is given by
$V= \Lambda^4 \left(1- \cos \left(\phi/f\right) \right).$
This model is controlled by the above discrete shift symmetry of the inflaton.
In natural inflation, it is known that 
natural inflation is consistent with 
the recent Planck results only for super-Planckian values of a decay constant, 
$f \gtrsim 7 M_P$ at $2\sigma$ level \cite{Ade:2015lrj}, where $M_P = 2.4 \times 10^{18}$ GeV is the Planck scale.
(Hereafter we adopt the Planck unit $M_P=1$ unless otherwise stated.)
Some extensions of the natural inflation are proposed 
with a focus on the weak gravity conjecture \cite{ArkaniHamed:2006dz,Brown:2015iha,Choi:2015aem,Kappl:2015esy}.
Among them, one of extensions of natural inflation is to prepare multiple sinusoidal potentials:
\begin{gather}
V = \sum_i \Lambda_i^4 \cos \left(\frac{\phi}{f_i} + \theta_i \right) +V_0.
\end{gather}
An overlapping of multiple sinusoidal potentials provides locally flatter regions 
where slow-roll inflation can successfully take place even though the decay constant takes a sub-Planckian value.
Such models with multiple sinusoidal potentials are called multi-natural inflation \cite{Czerny:2014wza, Czerny:2014xja, Czerny:2014qqa}.\footnote{
In bottom-up approaches, phenomenological aspects of such multiple sinusoidal potentials for an axion dark matter 
have been investigated in Refs.~\cite{Higaki:2016yqk,D'Amico:2016kqm,Jaeckel:2016qjp,Higaki:2016jjh,Daido:2016tsj}.
See also Refs.~\cite{Graham:2015cka,Choi:2015fiu,Kaplan:2015fuy} for relaxion models.
}

In this paper, we consider an origin of such multiple sinusoidal potentials and a relevant symmetry
from the viewpoint of string inspired supergravity (SUGRA).
In theories accompanying extra dimensions, compactification of them with small radii is required for realistic situations.
Since the extra dimension can have a periodicity, 
there will appear (quasi-)periodic functions under transformations of moduli relevant to
the extra dimensions in low energy scales. 
Such a periodic property can lead to a UV completion of so-called multi-natural inflation model.
We show an illustrating example in type II superstring theories, i.e., intersecting/magnetized D-brane scenarios compactified on extra dimensions containing the $T^2/{\mathbb Z}_2$ orbifold.
It is known that in toroidal compactification Yukawa couplings can be expressed by one of elliptic functions \cite{Cremades:2003qj, Cremades:2004wa}, which consist of multiple sinusoidal functions.\footnote{It is also known that gauge coupling constants are written by the theta functions \cite{Berg:2004ek}.}
Then, a Neveu--Schwartz (NS) axion plays an important role in driving cosmic inflation.
Such an axion possesses the shift symmetry 
which originates from the original gauge symmetry for the NSNS two-form gauge field coupled to fundamental strings, and
reflects on the periodicities of the extra dimensions after the compactification.\footnote{
A similar concept is studied also in Ref.~\cite{Kobayashi:2016mzg} with a focus on modular invariance of the Dedekind eta function
in a bottom-up approach.
} This can be understood also by using string dualities. NS axions in type IIA are known to be mapped into (a part of) complex structures in type IIB via T-duality (mirror symmetry).
In the F-theory, seven brane positions, which are coordinates on the (periodic) extra dimension, are treated as  
complex structures in an unified manner. Such axion shift symmetries can be related to periodicities or modular invariance in this sense.

This paper is organized as follows.
In Section 2, we review Jacobi's theta functions which consist of multiple cosine functions. 
In Section 3, we briefly review intersecting/magnetized D-brane scenarios in type II superstring theories,
and construct an inflationary model by using a dynamical supersymmetry (SUSY) breaking model in hidden sector.
In Section 4, we show theoretical predictions of tensor-to-scalar ratio $r$, spectral index $n_s$ and its running $\alpha_s$.
Section 5 is devoted to conclusions and discussions.

\section{Jacobi's theta function}
In this section, we review the Jacobi's theta function and its mathematical properties.
Subsequently, we discuss a relation between the Jacobi's theta function and multi-natural inflation.

The Jacobi's theta function is known as one of elliptic functions which possesses quasi-periodicities, 
and its definition is given by 
\begin{gather}
\vartheta
\begin{bmatrix}
a\\[3pt] b
\end{bmatrix}
(\nu, \tau) = \sum_{l =-\infty}^{\infty} e^{\pi i (a+l)^2 \tau} e^{2\pi i (a+l)(\nu + b)}.
\label{jacobi}
\end{gather}
Note that the Jacobi's theta function has four kinds of arguments, $a, b, \nu$ and $\tau$.
$a$ and $b$ take real values, and $\nu$ and $\tau$ are complex parameters.
It is easily found that this function only converges if ${\rm Im} \, \tau>0$, otherwise this function is not well-defined.
Also, the Jacobi's theta function is 
quasi-periodic
under the transformations of $a, b$ and $\nu$,
\begin{align}
\vartheta
\begin{bmatrix}
a+1\\[3pt] b
\end{bmatrix}
(\nu, \tau) &= \vartheta
\begin{bmatrix}
a\\[3pt] b
\end{bmatrix}
(\nu, \tau), \qquad
\vartheta
\begin{bmatrix}
a\\[3pt] b+1
\end{bmatrix}
(\nu, \tau) 
= e^{2\pi i a} \, \vartheta
\begin{bmatrix}
a\\[3pt] b
\end{bmatrix}
(\nu, \tau),
\end{align}
\begin{align}
\vartheta
\begin{bmatrix}
a\\[3pt] b
\end{bmatrix}
(\nu+1, \tau) &= e^{2\pi i a} \, \vartheta
\begin{bmatrix}
a\\[3pt] b
\end{bmatrix}
(\nu, \tau), 
\qquad
\vartheta
\begin{bmatrix}
a\\[3pt] b
\end{bmatrix}
(\nu+\tau, \tau) 
= e^{-2\pi i(b + \nu + \tau/2)} \, \vartheta
\begin{bmatrix}
a\\[3pt] b
\end{bmatrix}
(\nu, \tau).
\end{align}
Under the $SL(2,{\mathbb Z})$ transformation
\begin{align}
\tau \to \frac{a\tau +b}{c \tau + d},
\label{modtra}
\end{align}
where $ad-bc=1,~ a,b,c,d \in {\mathbb Z}$,
the theta function shows similar properties, for instance,
\begin{align}
\vartheta
\begin{bmatrix}
a\\[3pt]  b
\end{bmatrix}
(\nu, \tau+1) &= e^{-\pi i (a^2 -a)} \, \vartheta
\begin{bmatrix}
a\\[3pt] a + b -\frac{1}{2}
\end{bmatrix}
(\nu, \tau), 
\\
\vartheta
\begin{bmatrix}
a\\[3pt] b
\end{bmatrix}
\left(\frac{\nu}{\tau}, -\frac{1}{\tau} \right) 
&= (-i\tau)^{1/2} e^{2\pi i( \nu^2/2\tau + ab)} \, \vartheta
\begin{bmatrix}
b\\[3pt] -a
\end{bmatrix}
(\nu, \tau).
\end{align}
This is often called modular transformation of $\tau$. The shift of $\tau$ plays a role 
to a shift symmetry of an inflaton in our case.

As shown in Eq.~\eqref{jacobi},
the Jacobi's theta function consists of infinite summation of sinusoidal functions weighted by a complex Gaussian factor $e^{\pi i \tau}$.
In \rf{Higaki:2015kta}, the authors showed that the slow-roll inflation can take place with the potential given by 
\begin{gather}
V = \frac{\Lambda^4}{2}e^{-\pi i \tau} \left( 
\vartheta
\begin{bmatrix}
0\\[3pt] 0
\end{bmatrix}
(0, \tau) - \vartheta
\begin{bmatrix}
0\\[3pt] 0
\end{bmatrix}
\left(\frac\phi{2\pi f}, \tau \right) \right),
\label{eq:elliptic}
\end{gather}
where $\Lambda$ denotes an inflation scale and $\phi$ is an inflaton field.
The shape of the inflaton potential is shown in Figure~4 of \rf{Higaki:2015kta} and is symmetric under the shift of inflaton, $\phi \to \phi + 2\pi f$.
For large values of ${\rm Im} \, \tau$, natural inflation is realized, while for smaller values of ${\rm Im} \, \tau~(\gtrsim 1)$, hilltop inflation model \cite{Boubekeur:2005zm} is obtained. 
In the latter case, sub-leading cosine parts are enhanced and considerably contribute to the shape of the inflaton potential.
As a result, one finds
a quartic potential of the hilltop inflation, i.e.,
\begin{gather}
V \simeq V_0 - \lambda \phi^4,
\end{gather}
where $V_0$ and $\lambda$ are constants.
This potential predicts a small value of tensor-to-scalar ratio in a region consistent with the Planck 2015 results \cite{Ade:2015lrj}.
Since the above potential seems to be specific,
there are other possibilities, for instance, that Re\,$\tau$ becomes an inflaton candidate instead of Re\,$\nu$.
In the next section, we show an origin of the Jacobi's theta function based on intersecting/magnetized D-brane scenarios.

\section{Intersecting/magnetized D-branes}
In this section, we briefly review the framework of intersecting/magnetized D-brane models compactified on tori,
which are parts of the extra dimensions.
After a short review, we concretely provide an inflaton potential including compactification moduli fields.

\subsection{Bi-fundamental fields and Yukawa couplings}
We suppose that the ten-dimensional spacetime consists of 
four-dimensional spacetime and six-dimensional extra dimensions and also that
the latter is decomposed into a two-dimensional torus (or toroidal orbifold) and certain four-dimensional extra dimensions, 
e.g., $T^2 \times X_4$. Later, $T^2$ is replaced with toroidal orbifold $T^2/{\mathbb Z}_2$.
In the following part, we assume that the compactification preserves the four-dimensional ${\cal N}=1$ SUSY.
We try to concretely construct an inflationary model only in the intersecting brane picture, 
since the intersecting brane model based on type IIA superstring theory 
is connected to the magnetized brane model based on type IIB supserstring theory via T-duality \cite{Cremades:2003qj, Cremades:2004wa}.
In intersecting brane models, there is a remarkable mechanism of generation of chiral matters
which are bi-fundamental representations under the gauge groups realized on 
spacetime-filling D-branes. 
Intersections of two stacks of such D-branes can generate such matters, their degeneracies and Yukawa coupling constants among them.\footnote{See \rf{Ibanez:2012zz} for a review, and references therein.}
Hereafter, we will focus on spacetime-filling D6-branes.
The zero-modes of openstring NS and Ramond (R) sectors (bosons and fermions)
appear at points where a stack of $N_a$ D6$_a$-branes and a stack of $N_b$ D6$_b$-branes intersect each other.
Such zero-modes correspond to supermultiplets transformed as bi-fundamental representation of
$(N_a, \bar{N}_b)$ or $(\bar{N}_a, N_b)$ under $SU(N_a) \times SU(N_b)$ gauge group realized on the respective D-branes
in the four-dimensional SUSY effective theory.
In addition, the degeneracy of such bi-fundamental supermultiplets is given by the number of intersection points
between D6$_a$-brane and D6$_b$-brane, $I_{ab}$.
In the following part, we focus on a single $T^2$, on which 
D6-branes are wrapping on 1-cycles, for simple explanation.
D6-branes are wrapping also on 2-cycles on $X_4$, such that
${\cal N} =1$ SUSY is preserved and chirality is correctly generated.

Once any intersection numbers in the model are fixed, the number of bi-fundamental supermultiplets 
and their quantum numbers are also determined.
Then, Yukawa couplings among such three bi-fundamental supermultiplets can be analytically written by the Jacobi's theta function 
on $T^2$ \cite{Cremades:2003qj, Cremades:2004wa}, i.e.,
\begin{gather}
y^{(T^2)}_{ijk} \sim \vartheta
\begin{bmatrix}
\delta_{ijk} \\[3pt]
0
\end{bmatrix}
(\varphi, \kappa), \label{yukawa}
\end{gather}
where we neglected an overall factor and
three quantities dependent on compactification moduli fields are given by
\begin{align}
\delta_{ijk} &= \frac{i}{I_{ab}} + \frac{j}{I_{bc}} + \frac{k}{I_{ca}} + \frac{s}{d},\\
\varphi &= \frac{I_{ab} \theta_c + I_{ca} \theta_b + I_{bc}\theta_a}{d},\\
\kappa &= \frac{|I_{ab} I_{bc}I_{ca}|}{d^2} (B+iA).
\label{kappa}
\end{align}
This is a consequence of sums of world sheet instantons induced by windings of open strings stretching among intersecting D-branes.
Here, $\theta_X ~ (X=a, b, c)$ denote Wilson-line phase fields, $A$ is the torus area (the K\"ahler modulus) field\footnote{
$A$ is normalized by string length, hence this is dimensionless.
}
and $B$ is so-called NSNS axion field. 
These compactification moduli fields are candidates of an inflaton, however,
we will regard the $B$ axion as an inflaton as seen later.
Flavor indices $i, j$ and $k$ label each of matter contents, e.g., $i=0,1,2, ..., |I_{ab}|-1$.
Also, we define $d = {\rm gcd}\, (I_{ab}, I_{bc}, I_{ca})$, $s \equiv s(i,j,k) \in \mathbb{Z}$ is a linear function on integers $i,j$ and $k$. 
The form of the Yukawa couplings
reflects on the modular invariance of the toroidal model 
under the transformation (\ref{modtra}) with $\tau \equiv B+iA $ in a T-dualized picture of Ref.~\cite{Kobayashi:2016ovu}\footnote{
See Refs.~\cite{Ferrara:1990ei,Cvetic:1991qm,Derendinger:1991hq,LopesCardoso:1991ifk,Ibanez:1992hc}
for heterotic string cases.
}, because a linear combination $B+iA$ is mapped into the complex structure modulus on the torus via T-duality.
Hence, our model is expected to be modular invariant, i.e., symmetric under $B \to B+ 1$ 
(and changes of wavefunction basis on the extra dimensions),
unless any fields develop vacuum expectation values (VEVs).
This axion shift symmetry or modular invariance is a consequence of toroidal compactification with periodicities.

It should be noted that all of elements in Eq.~\eqref{yukawa} can not posses non-vanishing values.
Whether the Yukawa elements are vanishing or not depends on coupling selection rules which are characterized by extra dimensional topologies in string theories,
\begin{gather}
i+j+k \equiv 0 \qquad ({\rm mod}~d).
\label{selectionrule}
\end{gather}
In toroidal compactifications, it is known that selection rules lead to discrete flavor symmetries \cite{Abe:2009vi}.
After imposing selection rules, summing up all non-vanishing elements with respect to $s$ provides the general form of Yukawa couplings in intersecting D-brane models.
In the next subsection, we will see that an inflaton potential appears through Yukawa couplings in the presence of strong dynamics.
Indeed, such a situation can be realized in a model of dynamical SUSY breaking.

Finally, we comment on the orbifold extensions.
In the following, we will treat the intersecting D-branes wrapping on the fixed points of the toroidal orbifold $T^2/\mathbb{Z}_2$.
It is known that the D-branes wrapping on the rigid cycles can not move freely in toroidal compact space and 
opens string moduli are frozen.
A merit in considering the rigid cycles is that the position moduli associated with the D-brane positions 
are stabilized. The same holds for the Wilson line phases.
The allowed values of the Wilson line phases are summarized in the T-dual side \cite{Abe:2013bca}.
For simplicity, we assume all the Wilson line phases are vanishing in the following parts.
Since an extension to the non-vanishing cases is straightforward, we do not touch such cases in this paper.
Thus, the orbifold model building is more useful in reducing the extra degree of freedom of light fields
in the low energy effective theory.\footnote{
If open string moduli were alive, they would tend to have a slightly steeper slope of the potential than that of 
$B$ in our case. 
When the potential consists of combinations of the theta function,
we find that $\partial_B V \sim \partial_B \vartheta \sim \delta^2$ and 
$\partial_\theta V \sim \partial_\theta \vartheta \sim \delta$ with $\delta < 1$.
}
Note that the physical open string zero-modes on the rigid cycles can be expressed as linear combinations 
of the zero-modes appearing at the intersection points of the D-branes on the bulk of the original two-dimensional torus.
Accordingly, the Yukawa couplings on the rigid cycles are also given by linear combinations, as shown in the next subsection.

\subsection{Inflaton potential via IYIT mechanism}
\begin{figure}[h]
\centering
\hspace{25pt}
\begin{minipage}{0.49\hsize}
\centering
\includegraphics[clip, width=0.95\hsize]{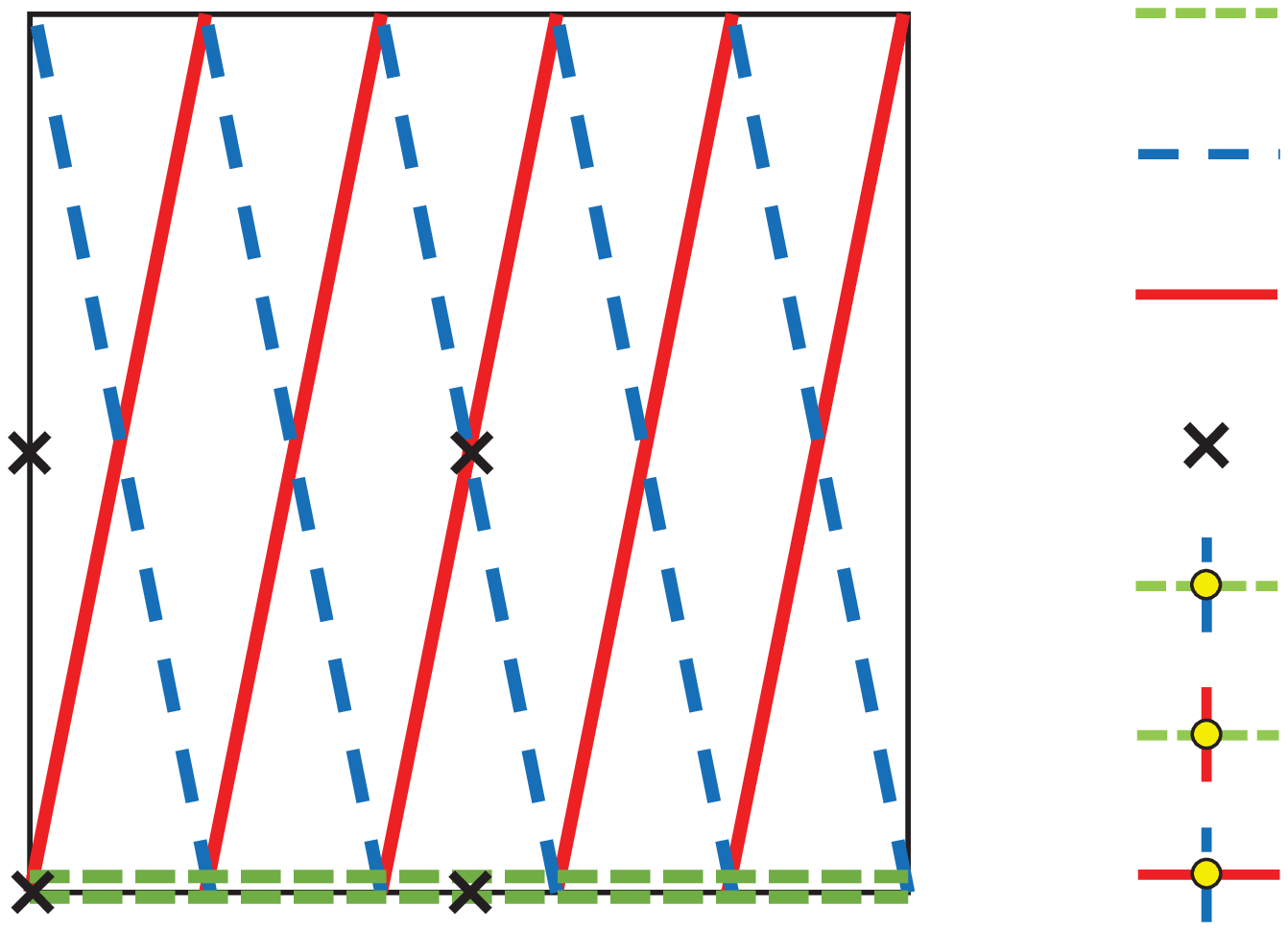}
\end{minipage}
\hspace{-40pt}
\begin{minipage}{0.49\hsize}
\centering
\vspace{-5pt}
\includegraphics[clip, width=0.70\hsize]{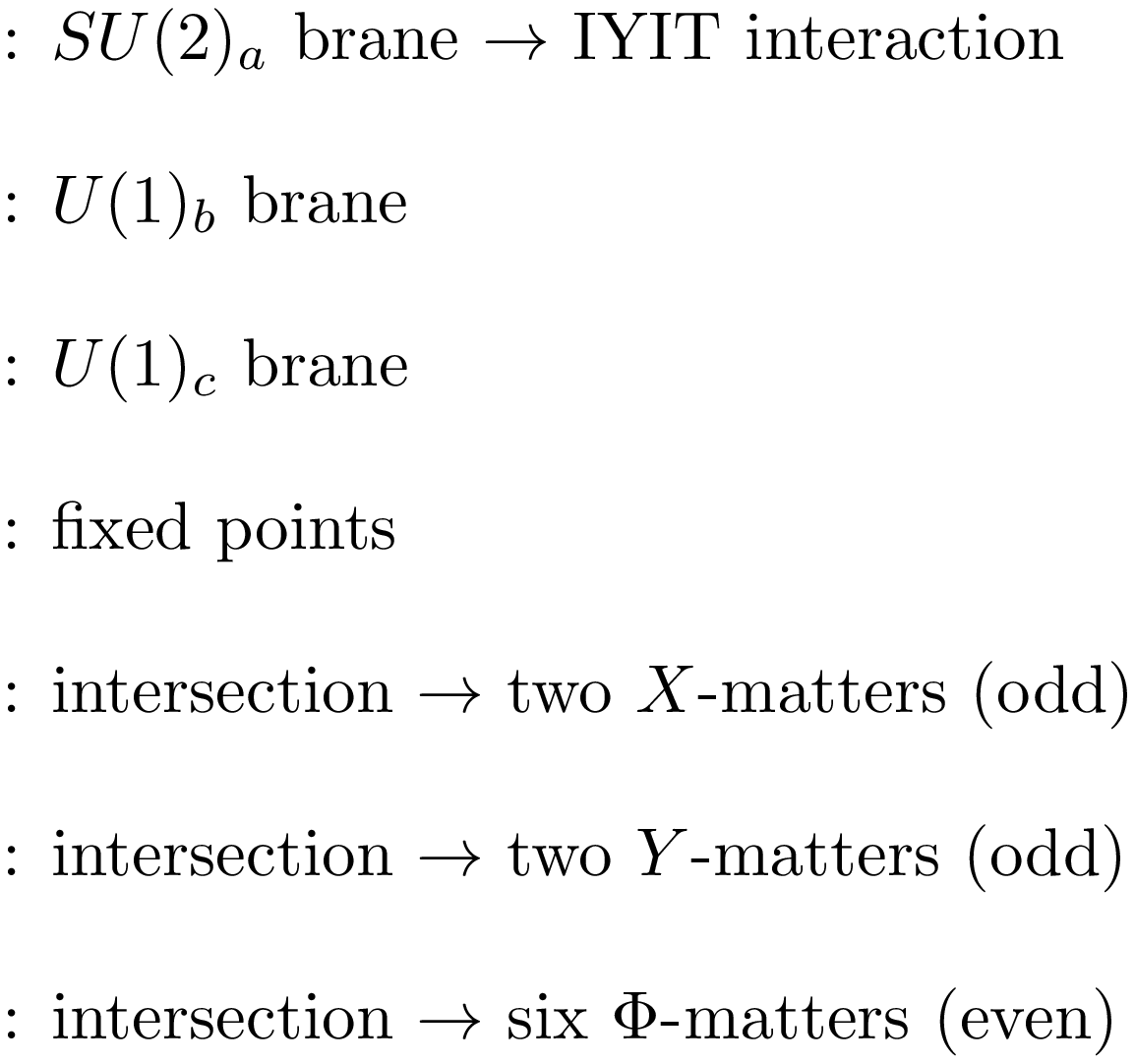}
\end{minipage}
\caption{The fundamental region of the torus and an example of the D-brane configuration.}
\label{branes}
\end{figure}

To obtain an inflationary potential energy, we use the Izawa--Yanagida--Intriligator--Thomas (IYIT) mechanism \cite{Izawa:1996pk, Intriligator:1996pu}. It is known as a model of dynamical SUSY breaking.
In the IYIT mechanism, four doublet and six singlet supermultiplets under $SU(2)$ gauge group are used to obtain a F-term scalar potential.
In our setup, we assume the following multiplicities of D-branes and their intersection numbers,
\begin{gather}
N_a=2, \qquad N_b=N_c=1,\\
|I_{ab}| = |I_{ca}|=5, \qquad I_{bc}=10. \label{intersection:iyit}
\end{gather}
We show an example of D-brane configuration in Figure~\ref{branes}.\footnote{
In this figure, we take $(n_a,m_a)=(1,0),~(n_b,m_b)=(1,-5)$ and $(n_c,m_c)=(1,5)$.
Then, one finds that
$I_{ab}^{(T^2)} = -5,~I_{bc}^{(T^2)} =10$ and $I_{ca}^{(T^2)} =-5$ only on $T^2$, where $I_{ab}^{(T^2)} = n_a m_b - m_a n_b $ and so on. 
To obtain Lorentz invariant Yukawa couplings, there must exist contributions of intersection numbers of 
$I_{ab}^{(X_4)}= \mp 1,~I_{bc}^{(X_4)}= \pm 1$ and $I_{ca}^{(X_4)} = \mp 1$
also from $X_4$. Here, total intersection number is given by, e.g., $I_{ab} = I_{ab}^{(T^2)} I_{ab}^{(X_4)}$.
Also, we may have other moduli contributions to Yukawa couplings from $X_4$,
and such moduli are neglected in this paper with an assumption that 
they are stabilized by, e.g., fluxes on $X_4$ or there are no contributions.
} 
In this model, the gauge symmetry $U(2)_a \times U(1)_b \times U(1)_c \simeq SU(2)_a \times U(1)_a \times U(1)_b \times U(1)_c$ is realized by the stacks of such D-branes. The indices denote gauge groups on corresponding D-brane, i.e.,
$U(2)_a$ gauge theory is living on the D$_a$-brane.
The $SU(2)_a$ gauge symmetry on the D$_a$-branes plays an important role in strong dynamics.
$U(1)$ groups are neglected below since some of them will become anomalous symmetry.
%
%
On the bulk of the original $T^2$, there are the five zero-modes for the $SU(2)_a$ doublets $| i \rangle_X \,\, (i=0,1, ..., 4)$ and 
$| j \rangle_Y \,\, (j=0,1, ..., 4)$. The former originates from intersection points betwen D$_a$- and D$_b$-branes,
and the latter comes from ones between betwen D$_c$- and D$_a$-branes.
In addition, there exist also the ten zero-modes for the 
$SU(2)_a$ singlets $|k\rangle_\Phi \,\, (k=0,1, ..., 9)$.
These originate from intersection points between D$_b$- and D$_c$-branes.
Here, $|0\rangle_X =|5\rangle_X$, $|0\rangle_Y=|5\rangle_Y$ and $|0\rangle_\Phi = |10 \rangle_\Phi$.
These zero-mode states transform under the $\mathbb{Z}_2$ orbifold projection as
\begin{gather}
\mathbb{Z}_2: \, |x\rangle \to |5-x \rangle \qquad \text{for  $X, Y$}, \qquad |y \rangle \to |10-y\rangle \qquad \text{for $\Phi$}.
\end{gather}
Then, we can find the physical states on the rigid cycles as
\begin{gather}
X_1= \frac1{\sqrt{2}} ( |1\rangle_X - |4\rangle_X ), \quad X_2=\frac1{\sqrt{2}}(|2\rangle_X - |3\rangle_X), \\
Y_3=\frac1{\sqrt{2}} ( |1\rangle_Y - |4\rangle_Y ), \quad Y_4= \frac1{\sqrt{2}}(|2\rangle_Y - |3\rangle_Y),
\end{gather}
for the $SU(2)$ doublets and
\begin{gather}
\Phi_1= |0\rangle_\Phi, \quad \Phi_2=\frac1{\sqrt{2}}(|1\rangle_\Phi + |9\rangle_\Phi), \quad  \Phi_3=\frac1{\sqrt{2}}(|2\rangle_\Phi +|8\rangle_\Phi), \notag\\
\Phi_4=\frac1{\sqrt{2}}(|3\rangle_\Phi + |7\rangle_\Phi), \quad \Phi_5=\frac1{\sqrt{2}}(|4\rangle_\Phi + |6\rangle_\Phi), \quad \Phi_6= |5\rangle_\Phi,
\end{gather}
for the $SU(2)$ singlets.\footnote{The detail of these physical states is written in Ref.~\cite{Abe:2008sx}.}
In final, we exhibit the matter fields in the IYIT mechanism in Table~\ref{matters}.

\begin{table}[h]
\centering
\begin{tabular}{|cccc|} \hline
Intersection & Degeneracy & Representation  & Field\\ \hline
$(a, b)$ & 2 & $\bm2_{(1,-1,0)}$ & $X_I \,\, (I=1,2)$\\
$(a, c)$ & 2 & $\bar{\bm2}_{(-1,0,1)}$ & $Y_J \,\, (J=3,4)$\\
$(b, c)$ & 6 & $\bm1_{(0,1,-1)}$ & $\Phi_K \,\, (K=1,2, ...,6)$\\ \hline
\end{tabular}
\caption{IYIT matter fields induced by the intersection numbers \eqref{intersection:iyit}.
The representation under $SU(2)_a\times U(1)_a \times U(1)_b \times U(1)_c$ is shown.}
\label{matters}
\end{table}

Then,
a superpotential among IYIT matters is given as
\begin{gather}
W= y_{IJK} X_I Y_J \Phi_K, \label{eq:iyit}
\end{gather}
where the degeneracies of IYIT matters are distinguished by flavor indices $I=1,2$, $J=3,4$ and $K=1,2, ... , 6$.
In the above superpotential \eqref{eq:iyit}, the Yukawa couplings $y_{IJK}$ on the rigid cycles can be analytically calculated by taking the linear combinations of the Yukawa couplings $y^{(T^2)}_{ijk}$ on the bulk in \eq{yukawa}.
For example, $y_{131}$ is written as
\begin{gather}
y_{131} = \frac12 (y^{(T^2)}_{110} - y^{(T^2)}_{410} - y^{(T^2)}_{140} + y^{(T^2)}_{440}).
\end{gather}
In the right hand side, four elements of the Yukawa couplings on the bulk are obtained by plugging each of flavor indices $i, j$ and $k$ into \eq{yukawa}.
Thus, by using the selection rules \eqref{selectionrule}, non-vanishing elements of the Yukawa couplings under consideration can be concretely written as
\begin{gather}
y_{131} = \sqrt{2}y_{145} = \sqrt{2}y_{235} = - (\eta_{10} + \eta_{40} + \eta_{60} + \eta_{90} + \eta_{110} ), \label{yukawa1}\\
y_{241} = -\sqrt{2} y_{143} = - \sqrt{2}y_{233} = - (\eta_{20} + \eta_{30} + \eta_{70} + \eta_{80} + \eta_{120} ),\label{yukawa2}\\
y_{136} = \sqrt{2}y_{142} = \sqrt{2}y_{232} = - (\eta_{15} + \eta_{35} + \eta_{65} + \eta_{85} + \eta_{115}),\label{yukawa3}\\
y_{246} = -\sqrt{2} y_{144} = -\sqrt{2} y_{234} = - (\eta_5 + \eta_{45} + \eta_{55} + \eta_{95} + \eta_{105}),\label{yukawa4}\\
y_{133} = y_{245} = \frac{1}{\sqrt{2}} (\eta_0 + 2\eta_{50} + 2 \eta_{100}),\label{yukawa5}\\
y_{242} = y_{134} = \frac{1}{\sqrt{2}} (2\eta_{25} + 2\eta_{75} + \eta_{125}),\label{yukawa6}
\end{gather}
where we define a shorthand notation,
\begin{gather}
\eta_N \equiv \vartheta
\begin{bmatrix}
N/250 \\[5pt]
0
\end{bmatrix}
(0, 10 (B+iA)).
\label{yukawa7}
\end{gather}
The other entries in the Yukawa couplings are all vanishing due to the selection rules in \eq{selectionrule}.\footnote{See Refs.~\cite{Abe:2008sx,Abe:2015yva} for details.}
As noted already, our model will be modular invariant at the current stage \cite{Kobayashi:2016ovu},
because no fields develop VEVs.
At the energy scale below a dynamically generated scale $\Lambda$, the low energy effective theory can be described by gauge-invariant meson fields $V_{IJ} \sim X_I Y_J$, and their effective superpotential is given by
\begin{gather}
W_{\rm eff} = y_{IJK} \Phi_K V_{IJ}, \label{eff}
\end{gather}
with
\begin{gather}
{\rm Pf} \, V_{ij} = \Lambda^4.
\end{gather}
The Pfaffian of the meson fields is expanded into the following form,
\begin{gather}
{\rm Pf} \, V_{ij} = V_{12} V_{34} - V_{13}V_{24} + V_{14} V_{23}.
\end{gather}
To evaluate the F-term scalar potential, we may choose a local minimum of three discrete local minima
\begin{gather}
{\rm Pf} \, V_{ij} =  V_{14}V_{23} = \Lambda^4,
\end{gather}
with the other mesons' VEVs vanishing.
Choosing the above local minimum, we rewrite the effective superpotential \eqref{eff} as
\begin{gather}
W_{\rm eff} = \Lambda^2 \sum_{K=1}^6 \left( y_{14K} \Phi_K +y_{23K} \Phi_K  \right).
\end{gather}
The lightest mode in the IYIT sector is SUSY breaking pseudo modulus, which is a linear combination of $\Phi_K$, and obtains the mass of order $\Lambda/4\pi$ at quantum level. Masses of other heavier modes are of order $\Lambda$. After integrating out them,
we consequently obtain the following F-term scalar potential,
\begin{gather}
V \simeq \sum_{K=1}^6 \left| \frac{\partial W_{\rm eff}}{\partial \Phi_K}\right| = \Lambda^4 \sum_{K=1}^6 | y_{14K} + y_{23K}|^2, \label{potential}
\end{gather}
where we set $|V_{13}| = |V_{24}| = \Lambda^2$, for simplicity.
Also, we neglect the supergravity effect and the VEVs of the $SU(2)$ singlets, $\langle \Phi_K \rangle \ll \Lambda$.
Note that the F-term scalar potential \eqref{potential} depends on the area of the torus $A$ and the NSNS axion $B$
in addition to $\Lambda$, 
because Yukawa couplings do on such moduli fields.
(See Eqs.~\eqref{kappa} and \eqref{yukawa1}\,--\,\eqref{yukawa7}.)
Although the above-mentioned calculation has been based on the intersecting D-branes in the IIA superstring,
we can eventually obtain the same scalar potential in T-dualized IIB superstring side
since the calculation will be equivalent to that in the magnetized D-branes.

At this stage, modular invariance will be broken down because meson fields develop VEVs,
if closed string moduli at the $SU(2)$ dynamical scale $\Lambda$ are stabilized. 
If not, there will exist modular invariance owing to a non-linear transformation of the moduli \cite{Kobayashi:2016ovu}, 
however, inflation is not viable then because the closed string moduli
disturb slow-roll condition. At any rate, we assume that the closed string moduli at $\Lambda$ is stabilized during the inflation
with flux compactifications \cite{Kallosh:2004yh,Abe:2007yb}, 
and hence modular invariance is broken down in the presence of meson VEVs as a consequence.
In other words, modular invariance of an original toroidal orbifold (in a T-dualized description) 
is broken down in the presence of D-branes. 
This is an analogue to an example that an original continuous shift symmetry of axion (inflaton)
is broken down to discrete one in the presence of instantons for natural inflation models.

Since the $B$-field appears as a dimensionless field, we should normalize it as 
\begin{gather}
B \to \phi/f
\end{gather}
with an axion decay constant $f$.
Here, $\phi$ is a canonically normalized axion.
$f$ may be expected to be of order of a compactification scale, because the NSNS $B$-field is a kind of closed string modulus 
field of toroidal compactifications.
Therefore, the scalar potential depends on $B,~A,~f$ and $\Lambda$.


\subsection{Gauge threshold corrections}
In the previous subsection, we obtained a scalar potential including an inflaton candidate of the $B$-field via 
Yukawa couplings with the IYIT mechanism.
It is known that there is the moduli dependence also in gauge couplings \cite{Berg:2004ek,Jockers:2004yj,Corvilain:2016kwe}.
Then, we include a gauge threshold correction to the dynamical scale $\Lambda$ in the $SU(2)$ gauge theory.

Hereafter, we take account of a one-loop correction to a gauge kinetic function on the stacks of specific D-branes, e.g., the stacks of $N$ D7-branes and $n$ D3-branes \cite{Berg:2004ek},\footnote{See also \rfs{Akerblom:2007uc, Akerblom:2007np}.}
\begin{gather}
f_\text{one-loop} \supset - \frac{n}{2\pi} \log 
\vartheta_1\left(\zeta, B+iA \right),
\qquad {\rm where}~~
\vartheta_1 (\zeta, B+iA) \equiv 
\vartheta
\begin{bmatrix}
1/2 \\[3pt] 1/2
\end{bmatrix}
\left(\zeta, B+iA \right).
\end{gather}
Here, $\zeta$ is a brane position modulus.
Further, there exists an another correction depending on the Dedekind's eta function $\eta(q)$
\begin{gather}
f_\text{one-loop} \supset - \frac{b}{2 \pi}\log \eta(q),
\end{gather}
where $b$ denotes a one-loop beta function coefficient in the ${\cal N}=2$ SUSY sector
and $\eta(q)$ is given by
\begin{gather}
\eta(q) \equiv q^\frac1{12} \prod_{n=1}^\infty (1-q^{2n}), \qquad {\rm where}~~q = \exp[i\pi (B+iA)] .
\end{gather}
Along with these corrections, the dynamical scale of the $SU(2)$ gauge theory is modified as
\begin{gather}
\Lambda^3 = e^{- \frac{2\pi}N f_\text{gauge}} \to \Lambda^3 \bigl[ \eta(q) \bigr]^{b/N} \bigl[ \vartheta_1(\zeta, q) \bigr]^{n/N}.
\end{gather}
Finally, we similarly recalculate the previous scalar potential \eqref{potential}, and then straightforwardly obtain a legitimate scalar potential,
\begin{gather}
V \simeq \Lambda^4 \hspace{1pt} | \eta(q) |^{4b/3N} \hspace{1pt} | \vartheta_1(\zeta, q) |^{4n/3N} \sum_{K=1}^6 | y_{14K} + y_{23K}|^2. \label{potential2}
\end{gather}
We numerically checked that these corrections from gauge coupling
do not essentially disturb the shape of the original potential \eqref{potential} in parameter regions of our interest
for $\zeta={\cal O}(1)$.\footnote{Note that the one-loop corrections to the dynamical scale are vanishing for the case of $\zeta=0$ for instance. 
This stems from the appearance of new massless modes.
However, as long as the effective theories are viable, such a situation does not take place.}
Thus, we will neglect the factors of $\eta$ and $\vartheta_1$ hereafter.
This is because the leading terms of $|\eta(q)| \simeq |q|^{1/12}$ and $|\vartheta_1(\zeta, B+iA)| \simeq |q|^{1/4}$ 
change just an overall scale,
and sub-leading terms of ${\cal O}(q^2)$ corrections in $\eta$ and $\vartheta_1$
do not drastically affect the inflaton potential.

\section{Numerical analysis}
In this section, we show the shape of the scalar potential in \eq{potential}
and inflationary predictions. 
As explained already, our model looks similar to hilltop inflation with a tuning of the torus area $A$.
Here and hereafter, we assume the moduli stabilization mechanism such that the area modulus $A$ 
is stabilized at a proper value of $A={\cal O}(1)$ and is heavier than the Hubble scale $\sim \Lambda^2$ during inflation.
This is expected due to the fact that the leading K\"ahler potential depends only on the modulus $A$ \cite{Blumenhagen:2006ci,Baumann:2014nda}, for instance,
\begin{gather}
K \sim - \log (T + \bar T), \label{kahler}\\
T \equiv A - i B.
\end{gather}
%
In type IIB superstring side, the modulus $T$ is connected to the complex structure modulus via T-duality.
Lots of complex structure modulus can be appropriately stabilized in the framework of flux compactifications and SUSY breaking mechanisms \cite{Giddings:2001yu,Kachru:2003aw,Berg:2005yu}.
Our scenario would be available if flux does not fix a complex structure in the absence of a specific direction of the flux and SUSY breaking fixes just a (real) part of it \cite{Higaki:2015kta,Hebecker:2015rya,Kobayashi:2015aaa}.
However, that is beyond the scope of this paper, and we leave a detail of moduli stabilizations for future work.

Now that the scalar potential in \eq{potential} 
is assumed to be a function of a dynamical $B$ with parameters of $A,~f$ and $\Lambda$.
With the K\"ahler potential of  Eq.~\eqref{kahler}, the axion decay constant $f$ has a connection with the torus area $A$ \cite{Baumann:2014nda}
\begin{gather}
f=\frac1{\sqrt2 A}.
\label{f-A}
\end{gather}
We canonically normalize the $B$ axion as $B \to \phi/f$ with this decay constant.
Also, a combination of $A$ and $\Lambda$ is fixed to fit the adiabatic 
density perturbation observed by the Planck satellite.
After all, we have only a single parameter, roughly speaking $A$, in the scalar potential with the dynamical $B$.


\subsection{The shape of an inflaton potential}
\begin{figure}[H]
\centering
\begin{minipage}[]{0.55\textwidth}
\centering
\includegraphics[width=\textwidth]{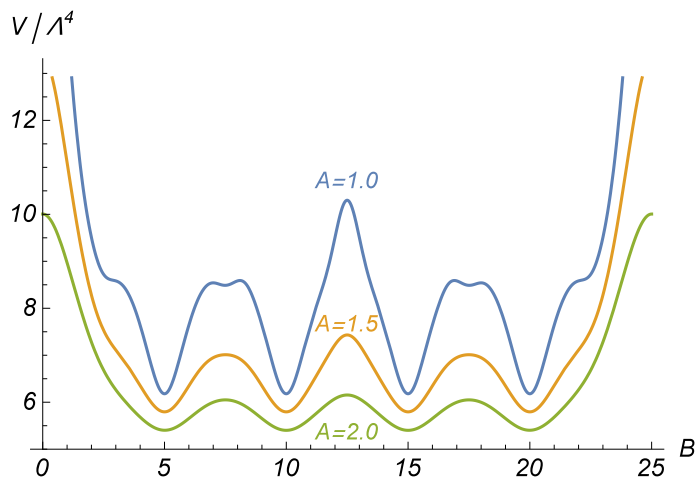}
\end{minipage}
\caption{Schematic pictures of the inflaton potential along the NSNS $B$-field direction. 
Reading from top to down, three
(blue-, yellow- and green-colored) lines denote the potential for $A=1.0, 1.5$ and $2.0$, respectively.
The periodicity seems to exist as $B \equiv B + 25$.}
\label{fig:pot}
\end{figure}

\begin{figure}[H]
\centering
\begin{minipage}[]{0.55\textwidth}
\centering
\includegraphics[width=\textwidth]{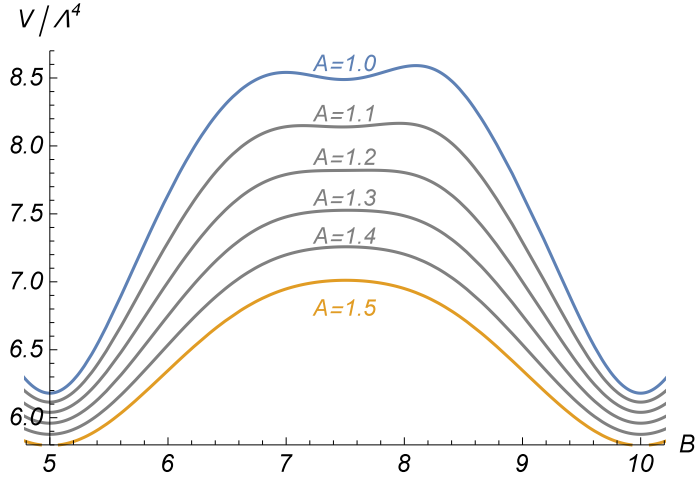}
\end{minipage}
\caption{The hilltop region of the inflaton potential. From top to down,
six lines correspond to $A=1.0, 1.1, 1.2, 1.3, 1.4$ and $1.5$.}
\label{fig:magni}
\end{figure}

First, the shape of the inflaton potential \eqref{potential} is discussed.
In Figure~\ref{fig:pot}, we show a schematic picture of the inflaton potential with varying values of the area of the torus $A$.
The blue, yellow and green lines correspond to the values of $A=1.0$, $1.5$ and $2.0$, respectively.
Figure~\ref{fig:pot} shows that there is a hilltop-like plateaux in the $B$-field direction around $B \simeq 7.5$ 
for $A \simeq 1.2$ (and we have the same structure around $B \simeq 17.5$). Similar studies are done also in Refs.~\cite{Higaki:2015kta,Kobayashi:2016mzg}.
Figure~\ref{fig:magni} is an enlarged view of the hilltop region in Figure~\ref{fig:pot}.
In the figure, the two bumps disappear for $A=1.2$ or more larger values.
From this figure, there seems to exist a periodicity of $B \equiv B + 25$. 
It is expected that D-branes wrapping on the torus change the original periodicity of $B \equiv B+1$
on the torus without D-branes.

\subsection{Relation to hilltop inflaton potential}
\begin{figure}[H]
\centering
\begin{minipage}[]{0.55\textwidth}
\centering
\includegraphics[width=\textwidth]{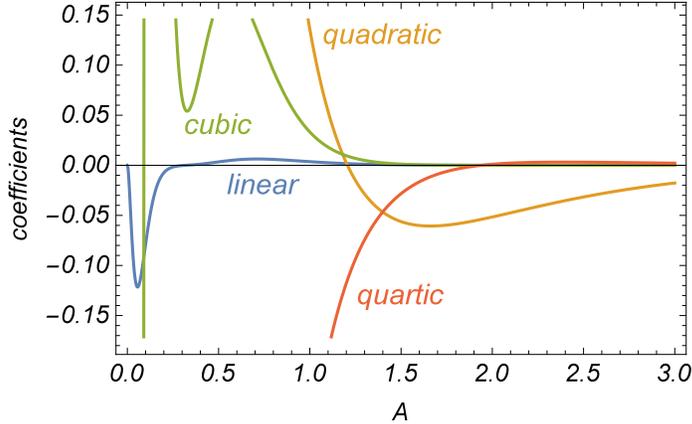}
\end{minipage}
\caption{$A$-dependence of $c_i$'s. We find $c_1 \simeq c_2 \simeq c_3 \simeq 0$ and $c_4 <0$ around $A \simeq 1.2$.}
\label{consideration}
\end{figure}

In this subsection, we explain how our model with theta functions can be interpreted as hilltop inflaton for $A \simeq 1.2$.
We expand the inflaton potential \eqref{potential} around the hilltop $B \simeq 7.5$, then the inflaton potential can be symbolically written by plugging $B=7.5 - \phi/f$ as
\begin{gather}
\frac{V}{\Lambda^4} = \frac{V_0}{\Lambda^4} + c_1(A) \frac{\phi}{f} + c_2(A) \left(\frac{\phi}{f}\right)^2 + c_3 (A) \left(\frac{\phi}{f}\right)^3 + c_4(A) \left(\frac{\phi}{f}\right)^4 + ...,
\end{gather}
where $V_0$ denotes a constant and $c_n$'s ($n=1,2,3,4$) are coefficients of the $\phi^n$-term 
depending on the modulus $A$.
The $A$-dependence of $c_n$'s are shown in Figure~\ref{consideration}.
We find that $c_1 \simeq c_2 \simeq c_3 \simeq 0$ and $c_4<0$ for $A \simeq 1.2$.
Then, the inflaton potential for $A \simeq 1.2$ can be expressed as
\begin{gather}
V \simeq V_0 - \lambda \phi^4,
\end{gather}
with $\lambda \propto c_4(A \simeq 1.2)$.
Note that this is not mere hilltop inflation but hilltop inflation with small linear- and cubic-terms. 
In Refs.~\cite{Takahashi:2013cxa,Harigaya:2013pla,Czerny:2014xja}, 
it is pointed out that such terms play an important role to give inflationary predictions different from hilltop inflation.

\subsection{Predictions of an inflation with $B$-field modulus}
In this subsection, we numerically analyze predictions on spectral index $n_s$, running of spectral index $\alpha_s$ 
and tensor-to-scalar ratio $r$. They are strictly constrained by the Planck 2015 results \cite{Ade:2015lrj},
and directly related to the slow-roll parameters defined by the scalar potential and its derivatives as below.
We find that these observables are consistent with the constraint in many choices of the decay constant $f$
smaller than the Planck scale in our model.\footnote{
See Refs.~\cite{Higaki:2015kta,Abe:2014xja} for models where a sizable $\alpha_s$ is realized in smaller $A$ cases.}

In a slow-roll inflation scenario with a single inflaton, $n_s$, $\alpha_s$ and $r$ are given by
\begin{align}
n_s &= 1 + 2\eta -6 \epsilon,\\
\alpha_s &= -24\epsilon^2 +16\epsilon\eta -2\xi,\\
r &= 16\epsilon.
\end{align}
Here, the slow-roll parameters are defined by
\begin{gather}
\epsilon = \frac12 \left(\frac{V'}{V}\right)^2, \qquad \eta = \frac{V''}{V}, \qquad \xi = \frac{V'V'''}{V^2}.
\end{gather}
A symbol $(')$ denotes a derivative with respect to $\phi$.

\begin{figure}[H]
\centering
\begin{minipage}[]{0.55\textwidth}
\centering
\includegraphics[width=1.02\textwidth]{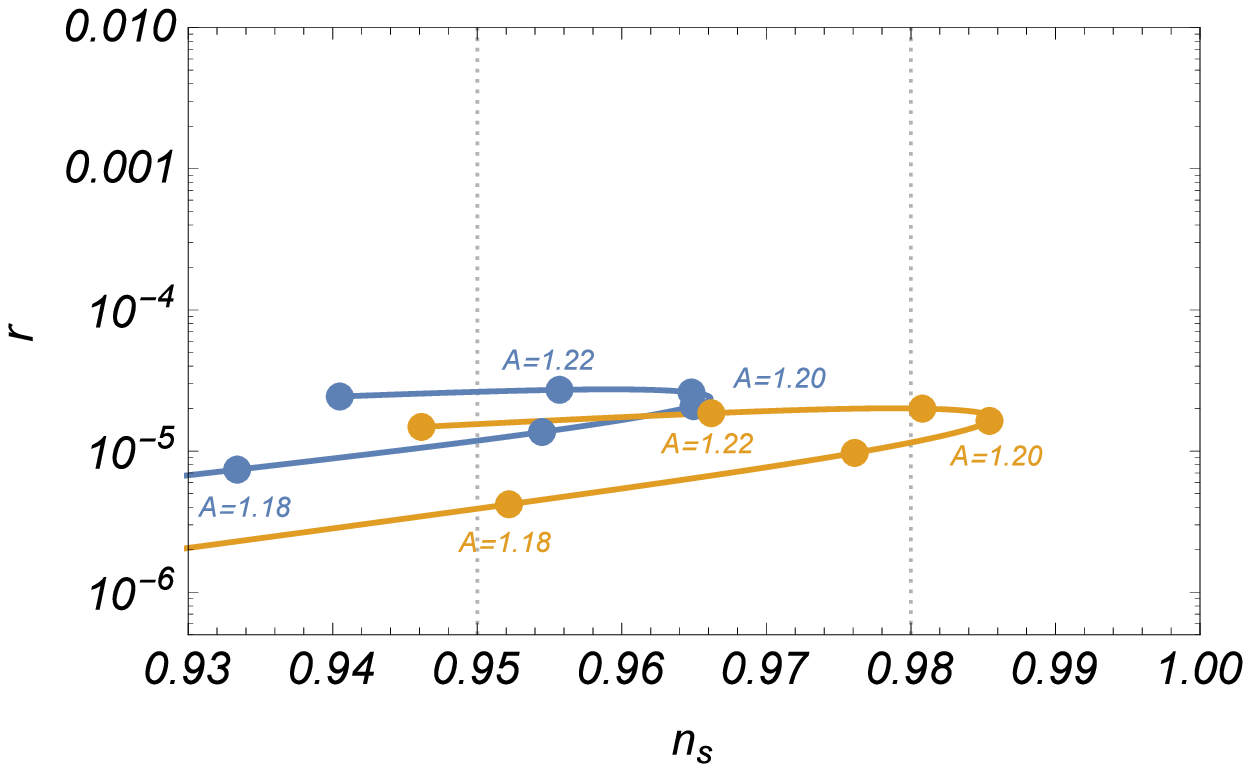}
\end{minipage}

\vspace{20pt}
\begin{minipage}[]{0.55\textwidth}
\centering
\includegraphics[width=\textwidth]{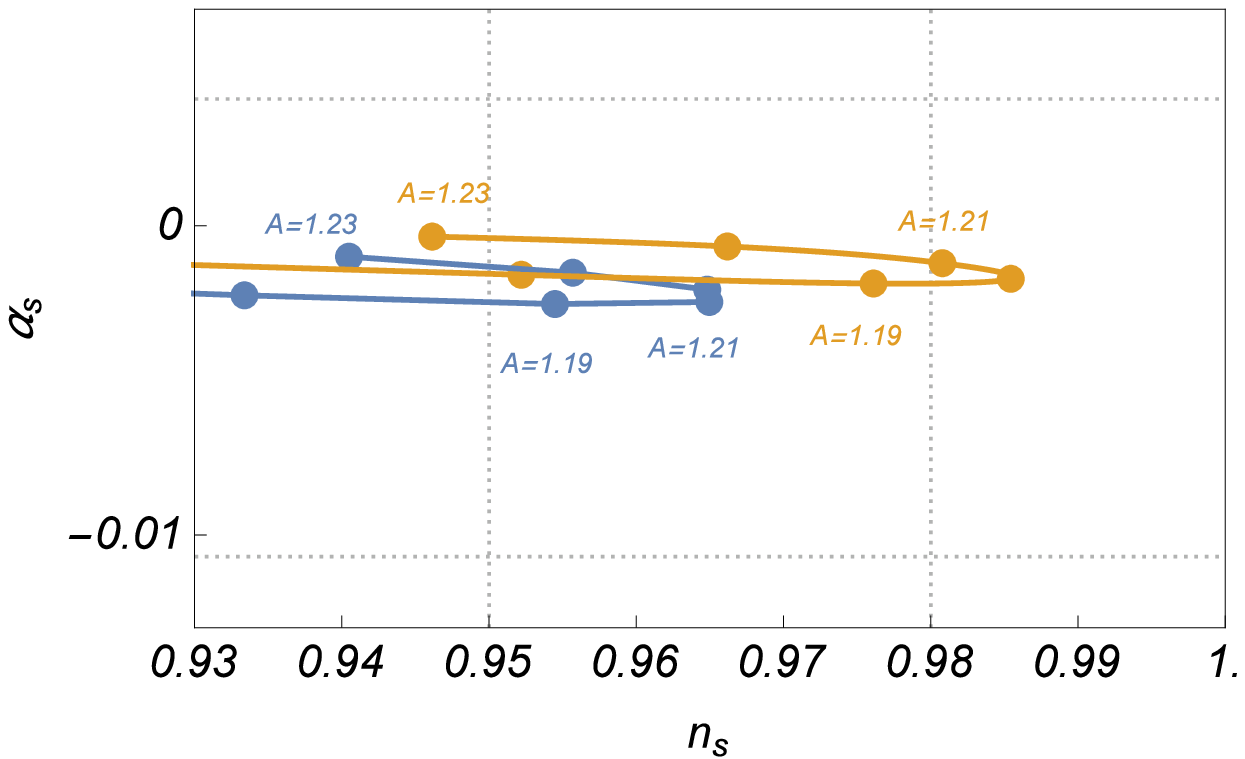}
\end{minipage}
\caption{Predictions on spectral index $n_s$, its running $\alpha_s$ and tensor-to-scalar ratio $r$. 
The yellow (blue) solid lines denote the e-foldings $N=60$ ($N=50$).}
\label{result}
\end{figure}

Figure \ref{result} shows results of observables ($n_s, r$ and $\alpha_s$) with numerical plots in the range of $1.17 \leq A \leq 1.23$.
The yellow (blue) curved lines denote the predictions for the e-foldings $N=60$ ($N=50$).
For a value of $A$ in this range, we obtain a sub-Planckian decay constant, $f \sim 0.58 < 1$, 
which leads to predictions consistent with 
the Planck 2015 observation:
\begin{gather}
r \sim 10^{-5}, \qquad \alpha_s \sim -0.001.
\end{gather}
Then, the inflation scale and inflaton mass $m_{\phi}$ are given by
\begin{align}
V_{\rm inf}^{1/4} & = 1.8 \times 10^{15}~{\rm GeV} \cdot \bigg(\frac{r}{10^{-5}}\bigg)^{1/4}  \sim \Lambda , \\
m_{\phi} &  \sim \frac{\Lambda^2}{f} \sim 2.3 \times 10^{12}~{\rm GeV} \cdot \bigg(\frac{r}{10^{-5}}\bigg)^{1/2}.
\end{align}
Since $\partial_{\Phi} W \sim \Lambda^2$ is the SUSY breaking scale,\footnote{
This IYIT sector may contribute to the vacuum energy as uplifting sector for obtaining Minkowski/de Siter vacuum from AdS one
in SUGRA \cite{Kachru:2003aw,Lebedev:2006qq,Dudas:2006gr,Abe:2006xp,Kallosh:2006dv,Abe:2007yb}, 
if $\Lambda^2$ is the main source of the SUSY breaking.
}
the gravitino mass is estimated as
\begin{align}
m_{3/2} \gtrsim \Lambda^2 \sim 1.3 \times 10^{12}~{\rm GeV} \cdot \bigg(\frac{r}{10^{-5}}\bigg)^{1/2} .
\end{align}
Here, we take into account the possibility that there may be other sources of the SUSY breaking.
Note that the Hubble scale during the inflation $H_{\rm inf}$ is given by $V_{\rm inf}^{1/2} \sim H_{\rm inf} \sim \Lambda^2$.
If reheating process takes place by the inflaton decay
through the axion-like interaction between $\phi$ and the Standard Model gauge bosons
\begin{align}
c \frac{g^2_{\rm SM}}{32 \pi^2} \frac{\phi}{f}  \epsilon^{\mu \nu \rho \sigma} F_{\mu \nu} F_{\rho \sigma},
\end{align}
the reheating temperature is estimated as
\begin{align}
T_R \sim 10^6 ~{\rm GeV} \cdot c \cdot \bigg(\frac{r}{10^{-5}}\bigg)^{3/4} .
\end{align}
Here, we used $4\pi/g_{\rm SM}^2 \sim 25$ for the Standard Model gauge coupling.
Several phenomenological consequences are similar to those shown in Ref.~\cite{Czerny:2014xja}.

\section{Conclusions and discussions}
In this paper, we have utilized the Jacobi's theta function to construct an inflationary model and investigated its predictions.
To this end, we have constructed a SUSY breaking model on $T^2/{\mathbb Z}_2$ and focused on the $B$-field (inflaton) dependence in the Yukawa couplings.
As a result, it is found that a hilltop-type inflation can be realized for a certain area of the torus ($A \simeq 1.2$).

We comment on corrections to the inflaton potential, and then the required conditions in our setup is discussed.
First, if there exist light modes with mass $\lesssim H_{\rm inf}$ during the inflation, the slow-roll inflation fails via mixings between 
the inflaton and extra fields. To evade this, 
the Hubble scale $H_{\rm inf} \sim \Lambda^2$ should be smaller than mass scales of possible extra modes. 
For instance, there is the pseudo modulus with mass $m_{\Phi} \sim \Lambda/4\pi$ and the other heavy modes with 
masses $\sim \Lambda$ in the SUSY breaking sector. Against these, the condition is naturally satisfied for $\Lambda < 1$.
Further, it is noted that string moduli stabilization can distort our inflaton potential generically, and vice versa.
If there exist moduli whose masses are smaller $H_{\rm inf}$ during the inflation,
vacuum of such moduli can be destabilized by the inflation energy of $3H_{\rm inf}^2$ 
and run away to decompactification vacuum, thus slow-roll inflation fails.
Hence, string moduli should be heavier than $H_{\rm inf}$.
Then, vacuum of string moduli can be stable even during the inflation. However, an inflaton potential can generally be made much steeper 
by large corrections from a scalar potential making moduli sufficiently heavy
\cite{Kachru:2003sx,Baumann:2007np} 
(when the stabilization potential depends on the inflaton).
A way to elude this issue is to decouple potential energy of heavy moduli 
from inflaton sector in a supersymmetric manner during the inflation.
This can be done by so-called strong moduli stabilization \cite{Kallosh:2004yh,Dudas:2012wi}, i.e., $W(\Phi) =\partial_{\Phi}W =0$ in the (inflationary) vacuum. Here, $W(\Phi)$ is a superpotential of heavy moduli $\Phi$.  
Then, our inflation model becomes viable, although such a stabilization may require a tuning of parameters in $W(\Phi)$ to cancel a steep part of the inflaton potential.\footnote{Additional conditions of moduli stabilization will be required
not to prevent a slow-roll inflation in the presence of the mixing between 
the inflaton and $\Phi$ in $W(\Phi)$ \cite{Kachru:2003sx,Baumann:2007np}.}
%
%
So far, we ignored SUGRA corrections.
Supposing the VEV of the pseudo modulus in the SUSY breaking sector $\langle \Phi_K \rangle \sim \Lambda^2$ \cite{Kitano:2006wz}, 
the inflaton potential is to be modified as $V_{\rm SUGRA} \simeq V + {\cal O}(\Lambda^6)$ owing to the shift symmetry of $B$.
Our analysis is quantitatively valid because of $\Lambda^2 \sim 10^{-6}$.
Finally, we discuss (non-)perturbative quantum corrections via shift symmetry breaking \cite{Berg:2005ja,Kobayashi:2015aaa}, 
which will induce the inflaton mass, and we parametrize it as $\Delta V \sim (c/16\pi^2) \Lambda^4 \phi^2$, 
where a coefficient constant $c$ can
contain the Yukawa coupling squared and unknown quantum gravity effects.
For $c \lesssim 1$, the inflationary predictions do not drastically change (see Figure \ref{consideration}) \cite{Higaki:2015kta}.


Before closing this section, we would like to comment on the CP violation.
It is pointed out that the CP symmetry among the quarks in the Standard Model is broken by non-vanishing values of an imaginary part of the complex structure modulus $\tau$ in type IIB side \cite{Kobayashi:2016qag}.
Since the complex structure modulus corresponds to the NSNS $B$-field via T-duality, the $B$-field axion is likely to play an important role in the CP violation.
Indeed, if the inflaton develops its non-vanishing VEV after inflation, the CP symmetry is broken down
and the value of a CP-violating phase is determined by the VEV, the intersecting numbers of D-branes in the quark sector and so forth.
The construction of a concrete model in the visible sector containing the SM quarks and leptons 
is attractive from the phenomenological point of view. Such a study is left for our future work.

\section*{Acknowledgment}
TH and YT would like to thank Tatsuo Kobayashi for valuable comments on the Yukawa couplings and their modular invariance.
Also, YT would like to thank Wilfried Buchm\"uller, Fuminobu Takahashi and Masahiro Ibe for valuable discussions and comments.
This work is supported by Grants-in-Aid for Scientific Research (No.~26247042 [TH] and No.~16J04612 [YT]) from the Ministry of Education, Culture, Sports, Science and Technology in Japan.

\bibliographystyle{hep}
\bibliography{references}
\end{document}